\documentclass[12pt]{article}
\begin{document}
\bigskip \bigskip \bigskip

\centerline{\large \bf On some local  properties of
Yang-Mills vacuum}
\bigskip \bigskip
\centerline{\bf E.I.Guendelman}
\medskip
\centerline{ Department of Physics}
\centerline{Ben Gurion University of the Negev}
\centerline{Beer-Sheva, 83105 Israel}
\medskip
\centerline{email: guendel@bgumail.bgu.ac.il}
\bigskip
\centerline{ \bf  Aleksey Nudelman}
\medskip
\centerline{Department of Physics}
\centerline{University of California}
\centerline{Santa Barbara, CA  93106-9530 U.S.A.}
\medskip
\centerline{email: anudel@physics.ucsb.edu}
\bigskip \bigskip
\begin{abstract}
In this paper we  use a non-covariant gauge which makes some of the 
YM features manifest which otherwise do not appear explicitly in the Landau gauge 
used by Nielsen and Olesen.We
show,in fact, that the unstable mode which makes a large contribution to
the vacuum energy is non-perturbative even classically.
 \end{abstract}
\medskip

	It has been almost 20 years since G.Savvidy, N.K.Nielsen and
P.Olesen
discovered 
an  unstable mode in the YM vacuum \cite{nielsen1}. Many interesting
developments have
been shaped
into attractive physical theories since that time but the nature of the
vacuum had not been
understood yet and remains one of the puzzles of  field and string
theories.

     A note about our notation is in order.We will use the (+ - - -)
metric.Three colors will be denoted by the upper indexes (1) (2) (3) for 
vector fields and by upper indexes 1 2 3 for ghost variables. We
will be using the following quantities throughout this paper:
\begin{eqnarray}
A_\mu = A^{(3)}_\mu \nonumber \\
W_\mu = \frac{ {A^{(1)}_\mu - iA^{(2)}_\mu}}{\sqrt{2}} \nonumber \\ 
D_\mu=\partial_\mu - igA_\mu
\end{eqnarray}
	Vacuum 
energy will be understood as a sum of the zero point energies of the linearized 
equations of motion:

\begin{equation}
D_\mu D^{\mu}W_\nu -D_\nu D_\mu W^{\mu} - 2igF_{\nu\mu}W^{\mu}=0
\label{lineq}  
\end{equation}
The non-covariant gauge condition represented as:
\begin{equation}
W_1 = 0
\label{axial}
\end{equation}
will be called axial gauge.

\paragraph
	\\The linearized equation of motion (\ref{lineq})
corresponds to the 
equation of motion of the massless vector particle coupled to the
electromagnetic potential. The original Yang-Mills Lagrangian is invariant
under
\begin{equation}
\vec{\tau}\vec{A'_\mu}=U(\vec{\theta})\vec{\tau}\vec{A_\mu}U(\vec{\theta})^{-1}
-\frac{i}{g}(\partial_\mu U(\vec{\theta}))U(\vec{\theta})^{-1}
\label{gaugetr}
\end{equation}
the gauge transformation, where
$U(\vec{\theta})=e^{-i\vec{\tau}\vec{\theta}}$
and
$2\vec{\tau}$ are the Pauli matrices. The infinitesimal gauge
transformations
\begin{eqnarray}
W'_\mu=W_\mu + iA_\mu \epsilon -i\theta W_\mu -\frac{\partial_\mu
\epsilon}{g} \nonumber \\
A'_\mu =A_\mu +i(W_\mu \epsilon^* - W_\mu ^* \epsilon)
-\frac{\partial_\mu \theta}{g}
\end{eqnarray}
are derived from eq \ref{gaugetr}
,where $\epsilon=\frac{\theta^1-i\theta^2}{\sqrt{2}}$ and
$\theta=\theta^3$. The linearized Lagrangian retains a residual $U(1)$
gauge invariance after the Landau or axial gauge conditions were imposed.
Thus we have $U(\vec{\theta})=U(\epsilon)U(\theta)$
and the Landau gauge becomes
\begin{equation}
U(\epsilon)=e^{-\frac{i}{\sqrt{2}}
\left(\matrix{0&\frac{1}{D^2 }D_\nu W^\nu \cr
	 \frac{1}{D^{*2}}(D_\nu W^\nu)^* &0\cr}\right)}
\end{equation}
while the axial gauge corresponds to
\begin{equation}
U(\epsilon)=e^{-{\frac{i}{\sqrt{2}}
\left( \matrix{0&\frac{g}{D_1}W_1 \cr
         \frac{g}{D^*_1}W^*_1 &0\cr}\right) }}.
\end{equation}
\paragraph
	\\While we are guaranteed by gauge invariance that the Landau and  
the axial gauge should correspond to the same physical entity there is
a feature of equation (\ref{lineq}) that becomes prominent only in  the
axial
gauge, making its study a worthy exercise.Solving  eq.(\ref{lineq}) 
in
the axial gauge with a constant chromomagnetic background 
\begin{equation}
A_2=gBx
\label{back}
\end{equation}
is almost as
simple as in the Landau gauge if one notices
that it is invariant under Lorentz transformations in the $z$
direction.Thus, we can find a suitable Lorentz frame where
$D_3 W_\mu =0$ by considering any particular solution with well
defined momentum in the third direction..The rest of the calculations are
most easily done in the Fourier space. Noting that the equation for $W_3$ 
\begin{equation}
\left ( -e^2 - \frac{d^2}{d\rho^2} + \rho^2 \right)W_3=0
\label{harmo}
\end{equation}
is that of  a simple harmonic oscillator which we can immediately solve 
for $e$
\begin{equation}
e^2=2n+1
\label{odnamo}
\end{equation}
where $n$ is a non-negative integer.The rest of the equations are: 
\begin{eqnarray}
\left(-\frac{d^2}{d\rho^2}+\rho^2 \right)W_0 -e\rho W_2 =0 \nonumber \\
e \frac{d}{d\rho}W_0 +\rho \frac{d}{d\rho}W_2 -W_2 =0 \nonumber \\
\left(e^2 + \frac{d^2}{d\rho^2}\right)W_2 + e\rho W_0 =0
\label{threq}
\end{eqnarray}
where 
\begin{eqnarray}
\rho=\frac{k_2+gBx}{\sqrt{gB}} \nonumber \\
e=\frac{E}{\sqrt{gB}}
\end{eqnarray}
and $E$, $k_2$ are Fourier energy and momentum in the $y$ direction
respectively.
The system (\ref{threq}) can  expressed as a third order
differential
equation  for $W_2$ which in turn can be  reduced to the second order
differential
equation
\begin{equation}
\frac{d^2}{d\rho^2}u -\frac{2}{\rho} \frac{d}{d\rho} u +(e^2
-\rho^2)u=0\label{otvet}
\end{equation}
by making 
the substitution
\begin{eqnarray}
u=\rho \frac{d}{d\rho}W_2-W_2 
\end{eqnarray}
Equation (\ref{otvet}) has the following solution
\begin{equation}
u(\rho)=\left(C_1
F\left(-\frac{e^2+1}{4},-\frac{1}{2},\rho^2\right)  +C_2\rho^3
F\left(-\frac{e^2-5}{4},\frac{5}{2},\rho^2\right)\right)e^{-\frac{\rho^2}{2}}
\label{solution}
\end{equation}
 where $C_1$ and $C_2$ are
arbitrary constants and $F$ is a confluent hypergeometric function.
We
expect
$u(\rho)$ to be a
bounded function.Thus, we obtain the energies  
\begin{eqnarray}
e^2=4n-1 \nonumber \\
e^2=4m+5
\label{dvemo}
\end{eqnarray}
where $n$ and $m$ are arbitrary non-negative integers terminating
the confluent hypergeometric series.\bigskip 

We now  note that  $e^2 = -1$
is a well-known  unstable mode \cite{nielsen1}.
By explicitly writing down the non-linear terms omitted in equation
(\ref{lineq}),
\begin{eqnarray}
\left( |W_2|^2 +|W_3|^2 \right)W_0 -\left(W_2^2 + W_3^2 \right)W_0^*
\nonumber \\
\left( |W_0|^2 -|W_3|^2 \right) W_2 - \left(W_0^2 - W_3^2 \right)W_2^*
\nonumber \\
 \left(|W_0|^2-|W_2|^2 \right)W_3 -\left( W_0^2-W_2^2 \right )W_3^*
\label{nonlin}
\end{eqnarray}
where we have used the axial gauge condition, equation (\ref{axial}),
we now find that our gauge choice reveals a rather peculiar aspect of this
mode.
From the equation (\ref{threq}) we can conclude that $W_0$ and $W_2$ have
the same phase factor for any mode except an unstable one.Thus we may 
find
a necessary and sufficient condition for (\ref{nonlin}) to vanish for a
simple case when solutions of different frequencies are not mixed. For
example, if
we consider a harmonic oscillator mode, eq. (\ref{harmo}), it will
trivially
put (\ref{nonlin}) to zero.It is interesting to note that the solution
of the system (\ref{threq}) in an arbitrary Lorentz frame will make
(\ref{nonlin}) vanish unless an unstable mode is present.This supports a 
widely held conjecture that an unstable mode is a possible part of a
non-perturbative phenomena, yet to be understood. To show that system
(\ref{threq}) will make
(\ref{nonlin}) vanish only in the absence of the tachyonic mode we rewrite
(\ref{nonlin}) as 
 \begin{eqnarray}
(b+c)u = (bv+cv) \nonumber \\
(a-c)v=(au-cv) \nonumber \\
(a-b)v=(au-bv)
\label{fun}
\end{eqnarray}
where $W_0=\sqrt{ua}$, $W_2=\sqrt{vb}$, $W_3=\sqrt{vc}$, $a$, $b$, $c$
are some positive real functions and $u$, $v$ are phase factors. And we
have also
used the fact that $W_0$, being Lorenz transformed back to an    arbitrary
$k_3$ frame, produces $W_3=\frac{k_3}{E}W_0$ with the phase
factor for $W_3$ always coinciding with that of $W_2$ . Since  system
(\ref{fun})
 has only one solution $u=v$ , only modes with real value of $E$ can
make (\ref{nonlin}) vanish.We find no analog of vanishing of the
non-linear
term in the Landau gauge, except for the scalar which contribution to the
vacuum energy is canceled by ghosts.This phenomena should not be
surprising since axial and Landau gauges are not proper gauge conditions
for non-linearized Yang-Mills.
	
	The vacuum energy (V.E.) can now be easily calculated.We have
three
energy modes: the two from eq.(\ref{odnamo}) and one from eq.(\ref{dvemo})
these should be
written
down in the arbitrary $k_3$ frame and summed over according the usual
prescription: we enclose the system in a box of volume V. In the interval
$dk_3$ there are $\frac{gBV}{4\pi^2} $ states with momentum $k_3$
\cite{nielsen2}
\begin{eqnarray}
V.E.= \frac{gBV}{4\pi^2}\int_{-\infty}^{+\infty} dk_3
\sum_{n=0}^{\infty}  (\sqrt{gB(2n+1)+k_3^2} \nonumber \\
+\sqrt{gB(4n+5)+k_3^2}+\sqrt{gB(4n-1)+k_3^2} )
 \end{eqnarray}
This sum is easily evaluated with the  use of a generalization of Salam
and
Strathdee formula \cite{salams}
\begin{equation}
\int_{-\infty}^{+\infty}{dk_3 \sum_{n=0}^{\infty}
\sqrt{gB(an+b)+k_3^2}}=\frac{gB}{2a}\left(\frac{b^2}{2} + \frac{a^2}{12}
-\frac{ab}{2} \right)\ln\frac{gB}{M^2}
\label{salam}
 \end{equation}

where $M$ is proportional to a cutoff of integration in Schwinger's proper
time formalism.Derivation of (\ref{salam}) is identical to that of Nielsen
and Olesen or Salam and Strathdee except for arbitrary coefficients
$a$,$b$.	
Let us note for completeness that the imaginary part of the vacuum
energy
in the axial gauge is 
\begin{equation}
Im \left[ \frac{gBV}{4\pi^2}\int_{-\infty}^{+\infty} dk_3 \sqrt{-gB
+k_3^2} \right]=-\frac{V g^2 B^2}{8 \pi}
\end{equation}
which is identical to that in Landau gauge considered by Nielsen and
Olesen.
The real part of the vacuum energy in the axial gauge
\begin{equation}
V.E.=\frac{11V\left(gB \right)^2}{48\pi^2}\ln\frac{gB}{M^2}
+ \frac{B^2}{2}.
\end{equation}
coincides, as expected,  with the Nielson-Olesen result in the Landau
gauge. In particular energy is minimized below zero, that is below
energy of the perturbative vacuum for $gB_{min}=M^2 e^{\frac{-24
\pi^2}{11 g^2}}$

	In conclusion, let us note that there is a very interesting
relation
between ghosts $\eta$ and modes of
$W_\mu$. The ghost Lagrangian in the Landau gauge:
\begin{equation}
L=\eta_+^* D^{\mu *}D_\mu^* \eta_+ +\eta_-^* D^\mu D_\mu
\eta_- 
\label{landaughost}
\end{equation}
Both terms in in the Lagrangian (\ref{landaughost}) give us 
Schr\"{o}dinger equations for a simple harmonic oscillator.These
are
the terms which cancel the  contribution of the two scalar modes $W_0$
and $W_3$
\cite{nielsen1} in the Landau
gauge.(Landau gauge condition $D_{\mu}W^{\mu}=0$ does not make these modes
not independent since they differ by an arbitrary zero energy mode:
$\phi_0 \sim e^{-\sqrt{2eH(n+1)}|x_3|-eHx_1^2/n!}$, where notation of
\cite{nielsen1} was used)
In the axial gauge the ghost Lagrangian
\begin{equation}
L=\eta^{1*}\partial_1 \eta^1 +\eta^{2*}\partial_1\eta^2
\end{equation}
makes no contribution to the vacuum energy.
\paragraph
	\\Though some progress was made in recent years towards resolution of
Nielsen - Olesen instability \cite{progress} we still lack a deep
understanding of the underlying physics. It is quite possible that the
resolution will come from the supersymmetric YM where the vacuum energy is
strictly zero and the gluinos form a stable but non trivial vacuum
\cite{matinyan}. Another possible solution may come from the non-trivial 
surface term in
the Lagrangian \cite{witten} making use of the `non - perturbative'
constant magnetic
background. The third possibility may come from the string theory if the
recent non-perturbative results \cite{d-branes} are to lead to the
discovery of the string
vacuum.

\subsection*{Acknowledgements}

We would like to thank Emil Nissimov, David Owen and Svetlana Pacheva for
discussions and interest in this work. AN is grateful to Joseph
Polchinski for an interesting discussion and to Smitha Vishveshwara for
proofreading. EIG thanks Sergei Matinian for encouraging correspondence.

 \end{document}